# Low-Field Regime of Magnon Transport in Yttrium Iron Garnet

Hossein Taghinejad,[1,2,*] Kohtaro Yamakawa,[1,3] Xiaoxi Huang,[4] Yuanqi Lyu,[1,3] Luke P. Cairns,[1,3] Ramamoorthy Ramesh,[1,3,4] James G. Analytis[1,2,3,*]

1. Department of Physics, University of California, Berkeley, CA, USA.
2. Kavli Energy NanoSciences Institute, University of California, Berkeley, CA, USA.
3. Materials Sciences Division, Lawrence Berkeley National Laboratory, Berkeley, CA, USA.
4. Department of Materials Science & Engineering, University of California, Berkeley, CA, USA.

*Corresponding Authors:
h.taghinejad@berkeley.edu
analytis@berkeley.edu

**ABSTRACT:** Diffusive propagation of spin waves and their quanta, magnons, in the archetypal magnetic insulator yttrium iron garnet (YIG) are the subject of a surge of research for low-power and low-loss data communication. However, operation under external magnetic fields reduces magnon diffusion length, attenuates the voltage amplitude at measurement terminals, and complicates the architecture of magnonic devices. Here, we explore the low-field and field-free regime of diffusive magnon transport in YIG films. We demonstrate that the field-induced suppression of magnon diffusion length can be fully inhibited only at the zero-field limit. Even a modest field of 10mT attenuates the non-local spin voltage by ~20% in a transport channel of ~ 1µm long. Using Stoner-Wohlfarth macrospin simulations, we reveal that an often overlooked, in-plane uniaxial anisotropy becomes the critical parameter governing the field-free operation of magnonic devices. We further demonstrate a tenfold enhancement in the effective field associated with the in-plane uniaxial anisotropy of YIG films at low temperatures—a key finding for field-free operation of magnonic devices under cryogenic conditions.

Spin waves, the collective excitations of magnetic materials, and their quanta—magnons—are currently the focus of intense research as potential alternatives to electronic charge for data communication. The excitement surrounding this paradigm shift stems from the wide range of possibilities that spin waves could unlock. First, spin waves offer a fundamental solution to the problem of heat generation in electronic devices, an issue that has become more critical since the sunset of the Dennard's scaling law. [1] Communicating data through spin waves of magnetic insulators eliminate the need for electronic charge movement during device operation, hence efficiently reducing the Joule heating issue encountered in charge-based electronics. Second, the wave nature of magnons opens up alternative computation paradigms, where the phase and frequency can be exploited to encode information. This capability is gaining momentum for non-Boolean computing, particularly in emerging machine-learning applications. Third, spin waves support high-speed operation, with frequencies ranging from gigahertz in dipolar magnons to the terahertz range in exchange magnons. [2-4] The recent report of coherent generation of THz waves could further fuel this direction. [5] These promising features have driven extensive research into the use of magnons for communication, processing, and storage of information, giving rise to the field of magnonics.

Ideal material candidates for magnonic devices typically meet three key criteria: (i) they should be electronically insulating to prevent parasitic effects from charge conduction, (ii) they must exhibit low magnetic damping to allow magnons to propagate over long distances, and (iii) they should support room-temperature operation. While the number of materials meeting these requirements is steadily growing, [6-9] yttrium iron garnet (YIG, $Y_3Fe_5O_{12}$) remains the gold standard in magnonics. YIG possesses the lowest magnetic damping of any known material, [10, 11] enabling robust magnon propagation over millimeter-scale distances [3, 12] with a giant spin conductivity. [13] Combined with its desired insulating electronic properties (bandgap ≈ 2.8eV) [14] and high Curie temperature ($T_C$ ≈ 560 K), [15] YIG has become the archetypal platform for magnonic research. Spin waves in YIG are traditionally excited inductively via the Orsted field generated by current-carrying wires fabricated on YIG surface. [16-18] While this method is straightforward for proof-of-concept studies, the extended footprint of the Oersted field hinders high-density device integration. An alternative approach relies on the integration of heavy metals like platinum (Pt) with YIG for localized excitation and detection of magnons by exploiting the spin Hall effect (SHE) and interfacial spin accumulation at the metal-YIG interface. [19] This method has gained considerable interest due to its scalability and potential for dense device integration.

In the linear regime, the localized nature of magnon generation via SHE in heavy metals leads to a *diffusive* propagation mode that is characterized by a characteristic magnon diffusion length. [19] Diffusive transport experiments are often conducted under the application of an external magnetic field, which poses several drawbacks. The application of the magnetic field reduces the magnon diffusion length in YIG, which, by extension, attenuates the amplitude of the voltage produced by travelling magnons at a non-local detector placed away from the injection site. This is a major challenge, as the low signal-amplitude is already a critical issue in spintronics, and applying magnetic fields worsens the problem. Cornelissen et al. studied this effect in medium-to-high magnetic fields (from 10mT to 1T), showing a severe attenuation of the detected voltage. [20] A similar effect is reported in other garnet films such as $Tm_3Fe_5O_{12}$. [9] It is also important to consider that high-quality YIG films are often grown on gadolinium gallium garnet (GGG, $Gd_3Ga_5O_{12}$) substrates. At high magnetic fields, the GGG substrate plays a non-trivial role in magnon transport within the YIG layer. Although GGG lacks exchange stiffness, at high fields it can contribute paramagnons to the hybrid YIG/GGG transport channel. [21] Additionally, at high fields, enhanced dipolar coupling between the GGG substrate and the YIG film can induce parasitic effects in YIG, [22] particularly at low temperatures where quantum magnonics is of interest. Application of large magnetic fields also complicate magnonic device design by significantly altering magnon dispersion, affecting key parameters such as the excitation energy gap, group velocity, and the equilibrium density of magnons. [23, 24] Finally, relying on external magnetic fields for device operation complicates the architecture of magnonic devices and increases power consumption. Therefore, the low-field, and ideally zero-field, operation is essential for unlocking the full potential of magnonic platforms.

Here, we investigate the non-local transport of diffusive magnons in YIG thin films at the ultra-low magnetic field regime, below 1 mT, and down to the zero-field limit. Our experiments reveal that the suppression of non-local voltage due to the magnetic field persists down to the zero-field limit, highlighting the critical importance of field-free operation of magnonic devices. In such an extremely low-field regime, we demonstrate that the weak, often overlooked in-plane uniaxial magnetic anisotropy of YIG becomes the key factor governing details of non-local magnon transport. To elucidate the role of this anisotropy for zero-field operation, we develop a Stoner-Wohlfarth (SW) macrospin model that not only quantifies YIG's in-plane uniaxial anisotropy but also allows conceiving device concepts based on the geometric twist between the device orientation and YIG's in-plane easy magnetic axis. The success of SW model in explaining our experimental data has important implications: in the low-field regime, *incoherent* spins waves can serve as a probe for *coherent* magnetization switching in YIG thin films. Combined with harmonic analysis of spin signals, this finding effectively positions non-local magnon transport platforms as magnetometers seamlessly embedded within magnonic devices, expanding their utility beyond conventional spintronic applications.

Figure 1(a) presents an optical image of a representative magnon transport device, consisting of two Pt wires with dimensions of 50 μm × 400 nm × 7 nm patterned on an 80 nm-thick YIG film. One Pt wire serves as the magnon injector, and the second wire acts as the detector, separated by a distance ~1 μm. As established in previous studies, [13, 19, 25] the application of a charge current to the Pt injector drives the propagation of magnons in the underlying YIG film via two distinct mechanisms. The first mechanism is the magnon-Seebeck effect, a thermally driven process. Here, the charge current induces Joule heating near the injector wire, creating a temperature gradient that propels magnons towards the detector. The second mechanism is an electronic excitation that leverages the spin Hall effect (SHE) within the Pt wire. The strong spin-orbit coupling in Pt generates a transverse spin current when a charge current is applied. This spin current flows towards the Pt – YIG interface, leading to spin accumulation. Through interfacial exchange coupling, this spin accumulation interacts with the local moments in YIG and serves as a reservoir of angular momentum for the excitation of magnons. As such, a spin-flip scattering (e.g., from $+\frac{\hbar}{2}$ to $-\frac{\hbar}{2}$) in Pt transfers one unit of angular momentum $(+\hbar)$ from the spin accumulation to the YIG film, leading to the out-of-equilibrium excitation of a magnon directly underneath the injector wire. Such a localized excitation creates a gradient in the magnon density, driving their diffusive propagation towards the detector wire.

At the detector, the propagating magnons are absorbed by conduction electrons in the Pt wire, resulting in spin accumulation at the Pt – YIG interface. This spin accumulation subsequently induces a measurable open-circuit voltage along the Pt wire via the inverse SHE. To differentiate between the two excitation mechanisms, we employ the lock-in detection technique and perform the harmonic analysis of the open-circuit voltage. Accordingly, driving an AC charge current at a frequency $f_0$ = 7.77 Hz into the injector wire generates two voltage components at the detector: a first harmonic component ($V^{1f}$, at frequency $f_0$) corresponding to the out-of-equilibrium magnons electronically excited via the SHE, and a second harmonic component ($V^{2f}$, at frequency $2f_0$) representing magnons thermally excited via the spin-Seebeck effect. [19, 25]

As depicted in Figure 1(a), our measurements are performed under an external magnetic field, $B_{ext}$, applied within the device plane. When $B_{ext}$ is sufficiently strong, YIG's magnetic moments ($M_{YIG}$) align with the external field. Thus, rotating $B_{ext}$ by an angle $\varphi_B$ rotates $M_{YIG}$ and generates characteristic $\cos^2(\varphi_B)$ and $\cos(\varphi_B)$ angular dependences for the first and second harmonic voltages at the detector. [19] In Figure 1(b), we have demonstrated such angular characteristics for several magnetic fields. A critical observation, however, is the scaling of the $V^{1f}$ and $V^{2f}$ voltages with $B_{ext}$. As shown in Figure 1(c), increasing the magnetic field suppresses the amplitude of the non-local voltage, particularly the first harmonic component. Notably, the $V^{1f}$ decreases by over 80% as the magnetic field approaches 0.5 T. This significant reduction is attributed, phenomenologically, to the reduced diffusion length of magnons at high magnetic fields. [20] In contrast, the suppression of the $V^{2f}$ signal is much weaker and onsets only at relatively larger fields compared to $V^{1f}$. This discrepancy stems from the direct heat diffusion towards the detector wire, and the *localized* generation of spin-Seebeck voltage underneath the detector. [20] Thus, the field-induced reduction of magnon diffusion length impacts the $V^{2f}$ voltage much less than the $V^{1f}$ component. The suppression of non-local voltages becomes more pronounced as the distance between the injector and detector wires increases (Figure S1, Supporting Information (SI)). These trends emphasize the need for exploring the magnon transport in YIG under lower magnetic fields. Otherwise, high field operation significantly diminishes signal amplitude, which impedes the use of magnons for long-distance data communication and efficient computing in next-generation electronic devices.

We next focus on magnon transport in the low-field regime ($B_{ext}$ < 5mT). We start with field-rotation experiments and study the angular characteristics of the non-local voltages under small fields. As shown in Figure 2(a, b), reducing $B_{ext}$ results in a progressive deviation of the $V^{1f}$ and $V^{2f}$ components from the characteristic $\cos^2(\varphi_B)$ and $\cos(\varphi_B)$ line-shapes, indicating that the azimuthal angle of the YIG magnetization ($\varphi_{YIG}$) no longer aligns with that of the external magnetic field (i.e., $\varphi_{YIG} \neq \varphi_B$). This deviation becomes more pronounced at smaller magnetic fields, particularly at $B_{ext}$ = 150 µT. At such low fields, the angular dependence of $V^{1f}$ and $V^{2f}$ voltages exhibit two abrupt jumps separated by 180º. Additionally, these jumps show hysteretic behaviors in the bidirectional scan of $\varphi_B$ between 0º and 360º. We attribute these observations to the competing effect of the in-plane magnetic anisotropy in the YIG film. When $B_{ext}$ becomes comparable to the anisotropy field, rotating the magnetic field across the hard magnetic axis triggers sudden magnetization switching from one quadrant to the next, which leads to abrupt jumps in the $V^{1f}$ and $V^{2f}$ signals. The significance of such an in-plane magnetic anisotropy is further confirmed from the different coercive fields required to saturate the non-local voltage when $B_{ext}$ is applied in different in-plane directions. For instance, a coercive field of approximately 150 µT is enough to saturate the $V^{2f}$ signal when $B_{ext}$ is applied at $\varphi_B$ = 0º (Figure 2c), whereas a coercive field exceeding 500 µT is required when $B_{ext}$ is applied at $\varphi_B$ = 90º (Figure 2d). Such direction-dependent coercivity leads to the observed distortion of $V^{1f}$ and $V^{2f}$ line-shapes, as shown in Figure 2(a, b).

To further elucidate the role of in-plane uniaxial magnetic anisotropy of YIG, we conduct macrospin simulations based on the SW model. Considering the coordinate system shown in Figure 3(a, b), the system's energy ($E$), normalized to the effective YIG volume ($V_{eff}$), can be described as:

$$(E/V_{eff}) = -M_{YIG}B_{ext}\cos(\varphi_{YIG} - \varphi_B) + K\sin^2(\varphi_{YIG} - \varphi_{EA}), \qquad (1)$$

where the first term represents the Zeeman energy, and the second term corresponds to the anisotropy energy. In Equation (1), $K$ and $\varphi_{EA}$ are, respectively, anisotropy energy and the direction of the magnetic easy axis relative to the [11$\bar{2}$] crystal axis of the YIG film. Figure 3(a) illustrates the crystal axes of the YIG film, highlighting [111] direction as normal to the film, and [11$\bar{2}$] and [1$\bar{1}$0] directions aligning with the edges of the YIG substrate. The Pt wires are aligned parallel to the [1$\bar{1}$0] crystal axis. As depicted in Figure 3(b), all angles in our macrospin simulations are measured relative to the [11$\bar{2}$] direction.

Within the described simulation framework, we calculate the system's energy as a function of $B_{ext}$, $\varphi_B$, $K$, $\varphi_{EA}$, and $\varphi_{YIG}$. By identifying local minima in the energy landscape, while accounting for the history of the applied magnetic field, we determine the equilibrium position of the YIG magnetization, $\varphi_{YIG}$. This process generates a set of $\varphi_{YIG}$ values that we use to simulate the experimental V$^{1f}$ and V$^{2f}$ voltages and, hence, calculate the $K$ and $\varphi_{EA}$ parameters. Figure 3(c) illustrates this process, where green circles mark the equilibrium position of the magnetization as $B_{ext}$ = 150 µT is rotated from $\varphi_B$ = 0° to $\varphi_B$ = 360° in 10° incremental steps. Figure 3(d, e) shows the best fits of the SW macrospin model to the first and second harmonic voltages, respectively, yielding $K$ = (16 ± 2) J/m³ and $\varphi_{EA}$ = 30° ± 5°.

Our simulations accurately capture the key experimental features, namely the angular dependences, abrupt voltage jumps, and the hysteretic behaviors of $V^{1f}$ and $V^{2f}$ signals observed during bidirectional field rotation (see, Figure S2, SI). Additionally, the simulations successfully reproduce the experimental hysteresis loops shown in Figure 2(c, d). These agreements validate that our model is suitable for analyzing the role of in-plane anisotropy of YIG in non-local magnon transport experiments within the low-field regime. Thus, we rely on the SW fitting of the experimental data to quantize the scaling of the $V^{1f}$ voltage as $B_{ext}$ approaches zero, while fully accounting for the role of the anisotropy field. As shown in the highlighted part of Figure 1(c), the $V^{1f}$ signal increases asymptotically, but noticeably, by more than 20% as the external field is reduced from 10 mT to 150 µT. This key finding reveals that the field-dependence of the magnon diffusion length persists even in the limit of vanishingly small magnetic fields, further highlighting the importance of zero-field operation. Field-free operation not only inhibits the loss of the signal amplitude but also significantly simplifies the architecture of magnonic devices by eliminating the need for supplying external magnetic fields, hence reducing the overall system complexity.

The in-plane magnetic anisotropy of YIG is critical for the field-free operation of magnonic devices. To further elaborate this point, we analyze three device configurations where Pt wires are geometrically twisted relative to the easy axis of the YIG film (Figure 4(a-c)). Using the macrospin model biased with the empirical values of $K = 16$ J/m$^3$ and $\varphi_{EA} = 30°$, we simulate the hysteresis behavior of $V^{1f}$ and $V^{2f}$ voltages in response to an applied magnetic field. In all cases, $B_{ext}$ is applied within the YIG plane and perpendicular to the Pt wires. Distinct characteristics are observed across the three configurations. In the A-configuration (Figure 4(a)), where the Pt wires are oriented perpendicular to the easy axis of YIG, the $V^{2f}$ signal exhibits a box-like hysteresis, while the $V^{1f}$ signal remains constant at its maximum value for all values of $B_{ext}$. In contrast, in the C-configuration (Figure 4(c)), where the Pt wires are aligned parallel to the YIG's easy axis, $V^{2f}$ and $V^{1f}$ signals exhibit linear and parabolic field dependencies, respectively, without any hysteresis. The intermediate case, the B-configuration (Figure 4(b)), with Pt wires twisted 45° relative to the easy axis, shows a hybrid behavior between these two extreme cases. We note that the device structure measured in Figure 2 closely resembles the B-type configuration (with a twist angle of ~ 30°). The successful simulation of the experimentally measured hysteresis loops for this device (solid lines in Figures 2(c, d)) further validates that the simulations presented in Figure 4 are reliable predictions of device performance in alternative designs.

A key parameter for the field-free operation is the amplitude of the $V^{1f}$ voltage in the remnant state (i.e., at $B_{ext} = 0$). In the A-configuration, the remnant $V^{1f}$ signal remains at its maximum value without any drop. In the B-configuration, the remnant $V^{1f}$ decreases by 50%, while in the C-configuration, the non-local signal drops to zero in the absence of an external field. This systematic change in the remnant $V^{1f}$ voltage with the device twist angle stems from the interfacial coupling between conduction electrons in Pt and local magnetic moments of the underlying YIG layer. As schematically depicted in Figure 4(d), once the Pt wires are fabricated, the SHE dictates a fixed polarization ($\vec{\sigma}$) for the spin accumulation at the Pt – YIG interface. Consequently, the efficiency of magnon injection and detection, $\eta_{inj,det}$, depends on the orientation of the YIG magnetization and can be phenomenologically described as:

$$\eta_{inj,det} = g(\vec{\sigma}.\vec{M}_{YIG}). \quad (2)$$

In the limit of $B_{ext} = 0$, the YIG magnetization aligns with the easy magnetic axis, that is $\vec{M_{YIG}} \parallel \vec{EA}$. Therefore, the twist angle between $\vec{\sigma}$ and the magnetic easy axis, $\Delta\varphi_{EA}$, becomes the sole parameter determining the $\eta_{\text{inj,det}}$ and, by extension, the amplitude of the non-local voltage. As plotted in Figure 4(e), Equation (2) yields an analytical expression for the coupling efficiency in the form of $\eta_{\text{inj,det}} \propto \cos(\Delta\varphi_{\text{EA}})$, leading to a $\cos^2(\Delta\varphi_{\text{EA}})$ modulation for the remnant value of the first harmonic signal ($V^{1f} \propto \eta_{\text{inj}} \times \eta_{\text{det}}$). Figure 4(e) also includes the remnant $V^{1f}$ values obtained from our macrospin simulation for the three device configurations studied in Figure 4(a-c), demonstrating agreement between our simulations and the analytical equation. Thus, we conclude that the $\vec{\sigma} \parallel \overrightarrow{EA}$ ($\vec{\sigma} \perp \overrightarrow{EA}$) condition in the A-configuration (C-configuration) leads to the full retention (loss) of the non-local voltage in the limit of $B_{\text{ext}} \to 0$. Further predications regarding the angular dependence of the first and second harmonic non-local signals for devices with the A-, B-, and C-configuration are shown in Figure S3, SI.

We note that the parameter $g$ in Equation (2) is a pre-factor accounting for the details of spin-charge interconversion. It includes contributions from the spin diffusion length and Hall angle of Pt wires and the Pt|YIG interfacial spin-mixing conductance. [19, 26] In the low-field regime, we treat $g$ as a constant parameter, an assumption that is supported by the field independence of the $V^{2f}$ voltage as previously discussed in Figure 1. Due to the localized nature of the spin-Seebeck effect, we can assume that $V^{2f} \propto \eta_{\text{det}} = g(\vec{\sigma} \cdot \vec{M}_{\text{YIG}})$. Up to $B_{\text{ext}} \approx 50$ mT, the $V^{2f}$ signal exhibits no significant field dependence, suggesting that the g-factor remains largely unaffected by the external field. In contrast to the $V^{2f}$ signal, the $V^{1f}$ component shows a persistent field dependence down to zero field, indicating that the scaling of $V^{1f}$ with $B_{\text{ext}}$ primarily stems from the influence of field on parameters related to the magnon transport within the YIG channel, rather than the $g$ factor, which is relevant only at the Pt-YIG interface.

We emphasize that the SW macrospin model describes coherent magnetization switching, where all spins switch direction nearly simultaneously. On the other hand, the SHE leads to incoherent excitation of out-of-equilibrium magnons in YIG films, essentially generating a broad spectrum of magnons with varying frequencies, phases, and linear momenta. Therefore, the success of SW model in explaining our magnon transport experiments has a significant implication: *incoherently* excited spins waves and their quanta, magnons, can effectively serve as proxies for monitoring *coherent* magnetization switching in YIG thin films. While both first and second harmonic signals can, in principle, provide such information, the $V^{2f}$ signal is particularly advantageous due to its localized nature and its robustness against magnon transport parameters. Given the relationship $V^{2f} \propto g(\sigma.M_{YIG})$, with fixed values of $g$ and $\sigma$, $V^{2f}$ reflects YIG's magnetization. As such, the second harmonic signal mimics the role of the classic vibrating sample magnetometry (VSM) technique, but over a smaller length-scale defined by the size and the footprint of the device under test. This approach complements the data obtained from VSM, which averages over the entire material volume, thereby masking fine details and heterogeneities within the sample. We note that the YIG film can be effectively treated as a monodomain over our device footprint, supported by extensive studies showing that magnetic domains in YIG films extend over hundreds of micrometers. [27] Thus, the $M_{YIG}$ probed via the $V^{2f}$ signal, near the detector wire, properly represents the magnetization state of the YIG across the entire magnonic transport channel, including near the injector wire. In other words, we can leverage the second harmonic component as an embedded magnetometer within the magnon transport channel to probe YIG's magnetic properties concurrently with, yet independent of, the transport phenomena revealed by the $V^{1f}$ harmonic. In the following, we showcase this capability by characterizing the temperature-dependence of YIG's in-plane anisotropy via analyzing $V^{2f}$ voltage at different temperatures.

Despite its relatively low value of 16 J/m$^3$ at 200K, the in-plane uniaxial anisotropy of YIG dictates key characteristics of magnon transport at low- and zero-field regime. This observation motivates further investigations into how uniaxial anisotropy energy scales at lower temperatures. To achieve this, we rely on the temperature dependence of the second harmonic signal. Figure 5(a) shows the response of the $V^{2f}$ signal to the external magnetic field at several temperatures ranging from 300K to 50K. Notable changes in the width ($\Delta B$) and height ($\Delta V^{2f}$) of the hysteresis loops are observed. From $\Delta B$ we can obtain the switching field $B_s = \Delta B/2$, the external magnetic field at which switching of YIG's magnetization to the opposite direction occurs. The $B_s$ value directly reflects the strength of YIG's uniaxial magnetic anisotropy because magnetization switching occurs when the Zeeman energy surpasses the anisotropy energy in Equation (1). The extracted values of $B_s$ and $\Delta V^{2f}$ at different temperatures are plotted in Figure 5(b). By fitting our SW macrospin simulations to the experimental data, we extract the anisotropy energy, $K$, and the effective anisotropy field, $B_K = \frac{K}{\mu_0 M_{YIG}}$. The temperature dependence of the YIG's saturation magnetization is accounted for (see, Figure S4, SI). As depicted in Figures 5(c) and 5(d), both $K$ and $B_K$ increase rapidly as temperature decreases, reaching saturation values of ~42 J/m$^3$ and ~220 µT, respectively, at temperatures below 100K. These values reflect more than 15- and 10-fold enhancement in the anisotropy energy and effective anisotropy field compared to the room temperature values of 2.7 J/m$^3$ and 20 µT, respectively. Such an order-of-magnitude increase in the anisotropy field underscores the importance of often overlooked in-plane uniaxial anisotropy of YIG films for low-temperature and low-field magnonic applications.

Our observation of uniaxial anisotropy aligns with previous studies on YIG thin films grown by pulsed-laser deposition (PLD) on (111)-GGG substrates. [27, 28] However, from a crystallographic standpoint, this finding is surprising. Ideally, the cubic structure of YIG favors threefold magnetocrystalline anisotropy. To the first order, the anisotropy energy of a cubic crystal can be expressed as $E_{\text{cubic}} \propto K_1(\alpha_1^2\alpha_2^2 + \alpha_1^2\alpha_3^2 + \alpha_2^2\alpha_3^2)$, where $\alpha_i$s are directional cosine functions. Defining the polar angle $\theta$ relative to the [111] axis and the azimuthal angle $\varphi$ relative to the [11$\bar{2}$] axis yields the following expression for the cubic anisotropy energy:

$$E_{\text{cubic}} \propto K_1\left(\frac{1}{4}\sin^4(\theta) + \frac{1}{3}\cos^4(\theta) + \frac{\sqrt{2}}{3}\sin^3(\theta)\cos(\theta)\cos(3\varphi)\right) \quad (3)$$

For the threefold anisotropy term (third term in Equation (3)) to appear, the magnetization must obtain an out-of-plane component, meaning that $\theta \neq \pi/2$. While this condition can be met in bulk YIG, the strong demagnetization field and easy-plane anisotropy field in thin films generally constrain magnetization to the YIG plane, thus effectively quenching the threefold in-plane anisotropy.

We attribute the uniaxial in-plane anisotropy of YIG films to extrinsic growth-induced effects. However, we believe the root cause of this effect is not due to sporadic factors such as material heterogeneities. Instead, it likely arises from a systematic, underlying mechanism intrinsic to the pulsed-laser deposition (PLD) of YIG films. Interestingly, the easy-axis orientation, $\varphi_{\text{EA}} = 30º \pm 5º$ relative to the [11$\bar{2}$], obtained from our non-local magnon transport experiments matches those extracted from magneto-optical Kerr effect (MOKE) [28] and spin magnetoresistance (SMR) [27] measurements. The unanimity of conclusions made via three different techniques (MOKE, SMR, and magnon transport) on independently grown PLD films reinforces our hypothesis of a systematic origin for the in-plane uniaxial anisotropy. The 30º offset from the $[11\bar{2}]$ direction aligns with the octahedral Fe–O bonds in the YIG lattice. Such octahedral Fe atoms are prone to re-sputtering during the PLD process, [29] creating vacancies that have been associated with the rhombohedral distortion and residual strain in PLD-grown YIG films on (111)-GGG substrates. [28, 30] Therefore, we speculate that this vacancy-induced strain contributes to the emergence of the in-plane uniaxial anisotropy observed in PLD-grown (111)-YIG films. Additionally, the temperature-dependence of the uniaxial anisotropy in Figure 5(c, d) closely mirrors the behavior of YIG's magneto-elastic coefficient, [31, 32] both showing a sharp increase as temperature drops from the room temperature and saturating below 100K. This correlation further implies strain-related effects as a key factor in the observed anisotropy. Nonetheless, a dedicated study is required to fully understand this mechanism.

In conclusion, we explored non-local propagation of diffusive magnons under the influence of external magnetic fields, demonstrating the continued attenuation of spin signals at measurement terminals down to the zero-field limit. This finding carries a significant implication; optimal signal amplitude in magnonic devices is achieved at zero-field conditions. This conclusion addresses a major challenge in spintronics, the signal amplification, which is a barrier currently limiting the widespread adoption of spintronic technologies. Our experiments further emphasized the essential role of in-plane uniaxial anisotropy in understanding the unconventional behavior of non-local magnon transport in the low-field regime. Through semi-empirical micromagnetic simulations, we demonstrated how this anisotropy can be leveraged to effectively design the coupling efficiency between spin reservoirs (i.e., Pt wires) and the magnon transport channel (i.e., YIG) thorough geometric twist in device orientation. This approach offers a practical pathway to designing complex magnonic devices without requiring external magnetic bias, thus advancing the development of efficient, field-free magnonic systems. We also presented evidence suggesting an extrinsic origin, likely growth-induced strain, as the underlying mechanism driving the observed uniaxial in-plane anisotropy in YIG thin films. While this anisotropy is subtle at room temperature, it becomes pronounced at lower temperatures. Our findings advocate for the strategic use of strain-engineering in garnet films to enable field-free magnonic operations. Potential approaches include applying epitaxial strain through substrate engineering and targeted elemental doping to tailor the magnetic properties effectively.

**Methods Section:**

**Growth of Thin Films.** YIG thin films were grown on (111)-oriented GGG substrates using reflection high-energy electron diffraction (RHEED)-assisted pulsed laser deposition technique. The deposition was carried out at a substrate temperature of 700 °C under an oxygen partial pressure of 100 mTorr. A chemically stoichiometric ceramic YIG target was employed. A KrF excimer laser ($\lambda$ = 248 nm) with an energy density of 1.5 J/cm² and a repetition rate of 5 Hz was used to ablate the target. After deposition, the films were cooled at a controlled rate of 10 °C/min under atmospheric oxygen conditions.

**Device Fabrication.** We employed two electron-beam lithography (EBL) steps to fabricate the devices on the YIG substrate. In the first EBL step, the Pt injector/detector wires (50 μm × 400 nm) are defined. Following the pattering, a 7 nm-thick Pt layer was deposited via RF sputtering at a base pressure better than $2\times10^{-6}$ Torr. Then, the sample underwent lift-off by soaking in acetone overnight. In the second EBL step, electrical leads to the Pt wires and contact pads were patterned. Subsequently, a 40nm/5nm Au/Ti bilayer was deposited in an e-beam evaporator at a base pressure better than $5\times10^{-6}$ Torr. The Ti layer serves as the adhesion layer. Prior to the Au/Ti deposition, a gentle Ar$^+$ ion milling was performed to ensure a clean interface between Au/Ti and the underlaying Pt layer. Poly(methyl methacrylate) (PMMA) was used as the electron resist in both EBL steps. The fabricated devices were wire bonded to a measurement puck for measurements.

**Non-local Transport Measurements.** Non-local magnon transport measurements were conducted using the lock-in detection technique. A 0.6 mA AC charge current at a frequency of $f_0$ = 7.77 Hz was applied to the Pt injector wire to excite magnons in the underlying YIG film. At the detector wire, two lock-in amplifiers were used for the simultaneous measurement of the first and second harmonic components at frequencies of $f_0$ and $2f_0$, respectively. The harmonic signals were found to be independent of the frequency used in our lock-in measurements (Figure S5, SI). All measurements were carried out under vacuum within a Quantum Design PPMS system. To study the angular dependence of the first and second harmonic signals, samples were rotated within a fixed magnetic field.

**Stoner – Wohlfarth Macrospin Simulation.** To obtain the equilibrium direction of the YIG magnetization from SW model, we calculate the first and second derivatives of Equation (1) with respect to $\varphi_{YIG}$. For each field condition ($B_{ext}$, $\varphi_B$), the equilibrium position is identified when both derivatives are positive. In our calculations, we consider the temperature dependence of YIG's saturation magnetization. To account for the slight misalignment of the substrate relative to the applied field direction, cause by sample loading on the measurement puck, we apply a small correction to the apparent $\varphi_B$ values.

**Acknowledgements:**

JGA and HT acknowledge financial support from the Bakar Institute through the Bakar Prize. HT acknowledges the financial support of the Kavli Energy NanoScience Institute (ENSI) through the Heising-Simons Postdoctoral Fellowship at the University of California, Berkeley.

**References:**

(1) Dennard, R. H.; Gaensslen, F. H.; Yu, H.-N.; Rideout, V. L.; Bassous, E.; LeBlanc, A. Design of Ion-Implanted MOSFET'S with Very Small Physical Dimensions. *IEEE J Solid-State Circuits* **1974**, *9* (5), 256-268.

(2) Mahmoud, A.; Ciubotaru, F.; Vanderveken, F.; Chumak, A. V.; Hamdioui, S.; Adelmann, C.; Cotofana, S. Introduction to Spin Wave Computing. *J Appl Phys* **2020**, *128* (16), 161101.

(3) Kajiwara, Y.; Harii, K.; Takahashi, S.; Ohe, J.; Uchida, K.; Mizuguchi, M.; Umezawa, H.; Kawai, H.; Ando, K.; Takanashi, K.; Maekawa, S.; Saitoh, E. Transmission of Electrical Signals by Spin-Wave Interconversion in a Magnetic Insulator. *Nature* **2010**, *464* (7286), 262–266.

(4) Serga, A. A.; Chumak, A. V.; Hillebrands, B. YIG Magnonics. *J Phys D Appl Phys* **2010**, *43* (26), 264002.

(5) Taghinejad, M.; Xia, C.; Hrton, M.; Lee, K.-T.; Kim, A. S.; Li, Q.; Guzelturk, B.; Kalousek, R.; Xu, F.; Cai, W.; Lindenberg, A. M.; Brongersma, M. L. Determining Hot-Carrier Transport Dynamics from Terahertz Emission. *Science* **2023**, *382* (6668), 299-305.

(6) Lebrun, R.; Ross, A.; Bender, S. A.; Qaiumzadeh, A.; Baldrati, L.; Cramer, J.; Brataas, A.; Duine, R. A.; Kläui, M. Tunable Long-Distance Spin Transport in a Crystalline Antiferromagnetic Iron Oxide. *Nature* **2018**, *561* (7722), 222–225.

(7) Yuan, W.; Zhu, Q.; Su, T.; Yao, Y.; Xing, W.; Chen, Y.; Ma, Y.; Lin, X.; Shi, J.; Shindou, R.; Xie, X. C.; Han, W. Experimental Signatures of Spin Superfluid Ground State in Canted Antiferromagnet $Cr_2O_3$ via Nonlocal Spin Transport. *Sci. Adv* **2018**, *4* (4),1098.

(8) Parsonnet, E.; Caretta, L.; Nagarajan, V.; Zhang, H.; Taghinejad, H.; Behera, P.; Huang, X.; Kavle, P.; Fernandez, A.; Nikonov, D.; Li, H.; Young, I.; Analytis, J.; Ramesh, R. Nonvolatile Electric Field Control of Thermal Magnons in the Absence of an Applied Magnetic Field. *Phys Rev Lett* **2022**, *129* (8), 087601.

(9) Gao, J.; Lambert, C. H.; Schlitz, R.; Fiebig, M.; Gambardella, P.; Vélez, S. Magnon Transport and Thermoelectric Effects in Ultrathin $Tm_3Fe_5O_{12}$/Pt Nonlocal Devices. *Phys Rev Res* **2022**, *4* (4), 043214.

(10) Cherepanov, V.; Kolokolov, I.; L'vov, V. The Saga of YIG: Spectra, Thermodynamics, Interaction and Relaxation of Magnons in a Complex Magnet. *Phys Reports* **1993**, *229* (3), 81-144.

(11) Glass, H. L.; Elliott, M. T. Attainment of the Intrinsic FMR Linewidth in Yttrium Iron Garnet Films Grown by Liquid Phase Epitaxy. *J Crys Growth* **1976**, *34* (2), 285-288.

(12) Maendl, S.; Stasinopoulos, I.; Grundler, D. Spin Waves with Large Decay Length and Few 100 nm Wavelengths in Thin Yttrium Iron Garnet Grown at the Wafer Scale. *Appl Phys Lett* **2017**, *111* (1), 012403.

(13) Wei, X. Y.; Santos, O. A.; Lusero, C. H. S.; Bauer, G. E. W.; Ben Youssef, J.; van Wees, B. J. Giant Magnon Spin Conductivity in Ultrathin Yttrium Iron Garnet Films. *Nat Mater* **2022**, *21* (12), 1352–1356.

(14) Thiery, N.; Naletov, V. V.; Vila, L.; Marty, A.; Brenac, A.; Jacquot, J. F.; De Loubens, G.; Viret, M.; Anane, A.; Cros, V.; Ben Youssef, J.; Beaulieu, N.; Demidov, V. E.; Divinskiy, B.; Demokritov, S. O.; Klein, O. Electrical Properties of Epitaxial Yttrium Iron Garnet Ultrathin Films at High Temperatures. *Phys Rev B* **2018**, *97* (6), 064422.

(15) Anderson, E. E. Molecular Field Model and the Magnetization of YIG. *Phys Rev* **1964**, *134* (6), A1581-A1585.

(16) Morrison, N.; Taghinejad, H.; Analytis, J.; Ma, E. Y. Coherent Spin Wave Excitation with Radio-Frequency Spin-Orbit Torque. *J Appl Phys* **2024**, *136* (11), 113901.

(17) Pirro, P.; Vasyuchka, V. I.; Serga, A. A.; Hillebrands, B. Advances in Coherent Magnonics. *Nature Reviews Materials*. **2021**, *6* (12), 1114–1135.

(18) Chumak, A. V.; Vasyuchka, V. I.; Serga, A. A.; Hillebrands, B. Magnon Spintronics. *Nat Phys* **2015**, *11*(6), 453–461.

(19) Cornelissen, L. J.; Liu, J.; Duine, R. A.; Youssef, J. Ben; Van Wees, B. J. Long-Distance Transport of Magnon Spin Information in a Magnetic Insulator at Room Temperature. *Nat Phys* **2015**, *11* (12), 1022–1026.

(20) Cornelissen, L. J.; Van Wees, B. J. Magnetic Field Dependence of the Magnon Spin Diffusion Length in the Magnetic Insulator Yttrium Iron Garnet. *Phys Rev B* **2016**, *93* (2), 020403.

(21) Oyanagi, K.; Takahashi, S.; Cornelissen, L. J.; Shan, J.; Daimon, S.; Kikkawa, T.; Bauer, G. E. W.; van Wees, B. J.; Saitoh, E. Spin Transport in Insulators without Exchange Stiffness. *Nat Commun* **2019**, *10* (1).

(22) Serha, R. O.; Voronov, A. A.; Schmoll, D.; Verba, R.; Levchenko, K. O.; Koraltan, S.; Davídková, K.; Budinská, B.; Wang, Q.; Dobrovolskiy, O. V.; Urbánek, M.; Lindner, M.; Reimann, T.; Dubs, C.; Gonzalez-Ballestero, C.; Abert, C.; Suess, D.; Bozhko, D. A.; Knauer, S.; Chumak, A. V. Magnetic Anisotropy and GGG Substrate Stray Field in YIG Films down to Millikelvin Temperatures. *npj Spintronics* **2024**, *2* (29).

(23) Jin, H.; Boona, S. R.; Yang, Z.; Myers, R. C.; Heremans, J. P. Effect of the Magnon Dispersion on the Longitudinal Spin Seebeck Effect in Yttrium Iron Garnets. *Phys Rev B Condens Matter Mater Phys* **2015**, *92* (5), 054436.

(24) Ritzmann, U.; Hinzke, D.; Kehlberger, A.; Guo, E. J.; Kläui, M.; Nowak, U. Magnetic Field Control of the Spin Seebeck Effect. *Phys Rev B Condens Matter Mater Phys* **2015**, *92* (17), 174411.

(25) Liu, J.; Cornelissen, L. J.; Shan, J.; Van Wees, B. J.; Kuschel, T. Nonlocal Magnon Spin Transport in Yttrium Iron Garnet with Tantalum and Platinum Spin Injection/Detection Electrodes. *J Phys D Appl Phys* **2018**, *51* (22), 224005.

(26) Qiu, Z.; Ando, K.; Uchida, K.; Kajiwara, Y.; Takahashi, R.; Nakayama, H.; An, T.; Fujikawa, Y.; Saitoh, E. Spin Mixing Conductance at a Well-Controlled Platinum/Yttrium Iron Garnet Interface. *Appl Phys Lett* **2013**, *103* (9), 092404.

(27) Mendil, J.; Trassin, M.; Bu, Q.; Schaab, J.; Baumgartner, M.; Murer, C.; Dao, P. T.; Vijayakumar, J.; Bracher, D.; Bouillet, C.; Vaz, C. A. F.; Fiebig, M.; Gambardella, P. Magnetic Properties and Domain Structure of Ultrathin Yttrium Iron Garnet/Pt Bilayers. *Phys Rev Mater* **2019**, *3* (3), 034403.

(28) Sokolov, N. S.; Fedorov, V. V.; Korovin, A. M.; Suturin, S. M.; Baranov, D. A.; Gastev, S. V.; Krichevtsov, B. B.; Maksimova, K. Y.; Grunin, A. I.; Bursian, V. E.; Lutsev, L. V.; Tabuchi, M. Thin Yttrium Iron Garnet Films Grown by Pulsed Laser Deposition: Crystal Structure, Static, and Dynamic Magnetic Properties. *J Appl Phys* **2016**, *119* (2), 023903.

(29) Manuilov, S. A.; Khartsev, S. I.; Grishin, A. M. Pulsed Laser Deposited $Y_3Fe_5O_{12}$ Films: Nature of Magnetic Anisotropy I. *J Appl Phys* **2009**, *106* (12), 123917.

(30) Manuilov, S. A.; Fors, R.; Khartsev, S. I.; Grishin, A. M. Submicron $Y_3Fe_5O_{12}$ Film Magnetostatic Wave Band Pass Filters. *J Appl Phys* **2009**, *105* (3), 033917.

(31) Callen, H. B.; Callens, E. The Present Status of the Temperature Dependence of Magnetocrystalline Anisotropy, and the l(l+1)2 Power Law. *J Phys Chem Solid* **1966**, *27* (8), 1271-1285.

(32) Callen K, R. A.; Callen, H. B. Magnetostriction in Cubic Neel Ferrimagnets, with Application to YIG. Phys. Rev. **1963**, *130* (5), 1735-1740.

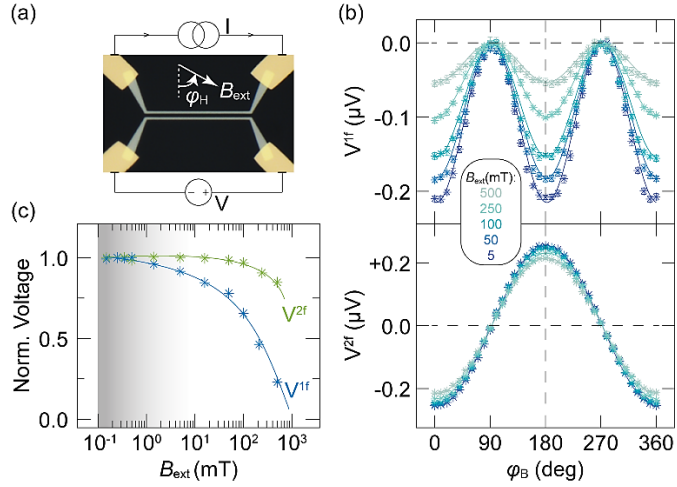

**Figure 1: Attenuation of Non-Local Magnon Signal With Magnetic Field.** (a) Optical image of a representative non-local magnon transport device. Two Pt wires spaced 1 μm apart are used for injection (top wire) and detection (bottom wire) of magnons in the underlying YIG film (~80 nm). Measurements are taken under an in-plane magnetic field ($B_{ext}$) applied at various angles ($\varphi_B$). (b) First (top) and second (bottom) harmonic voltages measured in response to rotating magnetic fields of varying intensities. Solid lines represent $V^{1f} \times \cos^2(\varphi_B)$ and $V^{2f} \times \cos(\varphi_B)$ fits to the respective harmonic signals. (c) Scaling of $V^{1f}$ and $V^{2f}$ signals with the magnetic field intensity. $V^{1f}$ and $V^{2f}$ signals at different magnetic fields are normalized to their respective maximum values within each dataset. Solid lines are guides to the eyes. All measurements are performed at 200 K.

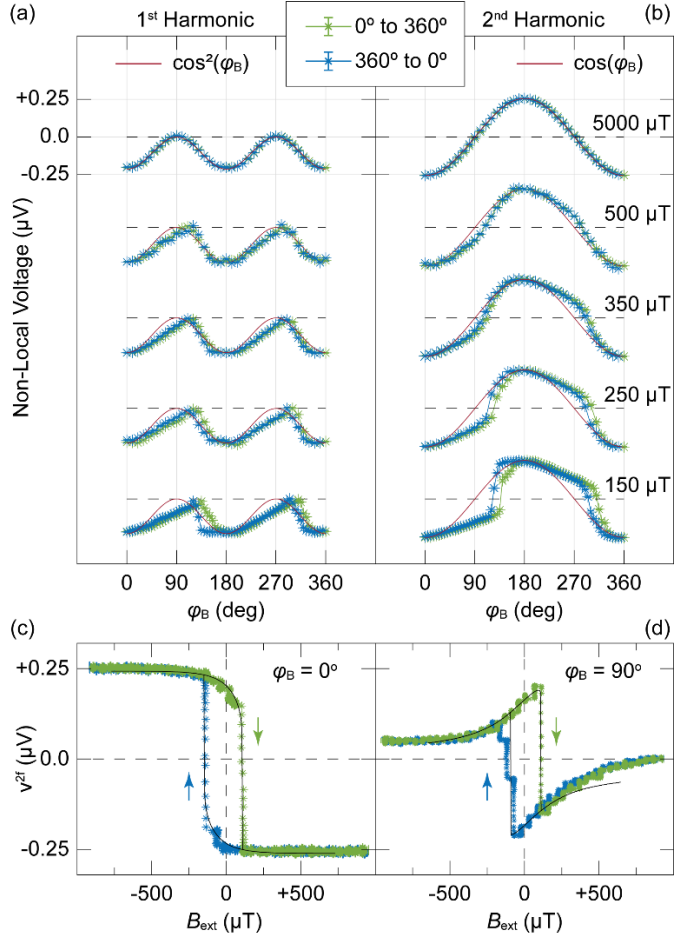

**Figure 2: Magnon Transport Under *Low* Magnetic Fields.** (a, b) Angular dependence of the first and second harmonic voltages, respectively, measured under low magnetic fields. For clarity, collected data at different magnetic fields are shifted vertically. Deviation from characteristic $\cos^2(\varphi_B)$ and $\cos(\varphi_B)$ angular dependences and pronounced hysteresis are observed at low magnetic fields. (c, d) Hysteresis loops of the second harmonic voltage, $V^{2f}$, with the magnetic field applied normal (i.e., $\varphi_B = 0°$) and parallel (i.e., $\varphi_B = 90°$) to Pt wires, respectively. Black lines represent the SW macrospin simulations, as explained in the text. Arrows indicate the direction of the magnetic field sweep.

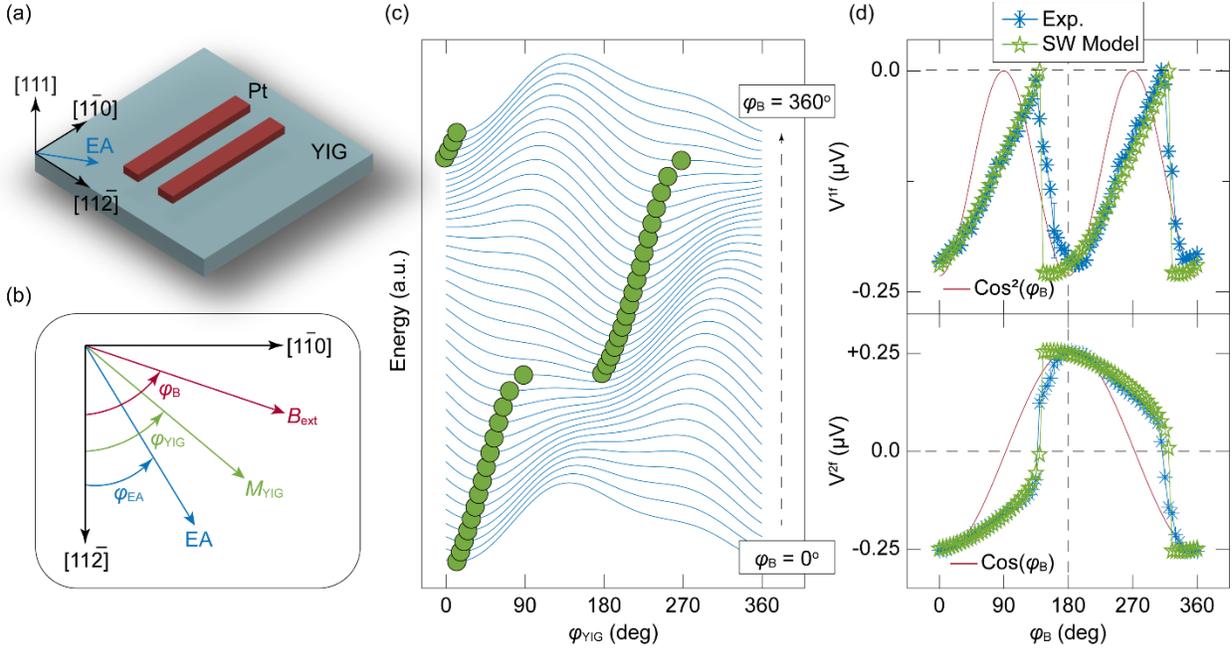

**Figure 3: Stoner – Wohlfarth (SW) Macrospin Simulations.** (a) Schematic illustration of the device geometry relative to the in-plane magnetic easy axis (EA) and crystal axes of the YIG film. (b) Modeling parameters within the $[11\bar{2}] - [1\bar{1}0]$ plane, including intensity and angle of external magnetic field ($B_{ext}$ and $\varphi_B$), equilibrium direction of YIG magnetization ($M_{YIG}$, $\varphi_{YIG}$), and the strength and direction of the uniaxial in-plane magnetic anisotropy ($K$, $\varphi_{EA}$). (c) Energy landscapes obtained from Equation (1) when $B_{ext} = 150$ µT is applied along $\varphi_B = 0°\text{-}360°$ (in $10°$ steps). The equilibrium direction of the YIG's magnetization is identified as the metastable energy minima, marked by green circles. (d) Simulating the first and second harmonic voltages using the SW model yields $K = 16 \pm 2$ J/m³ and $\varphi_{EA} = 30° \pm 5°$.

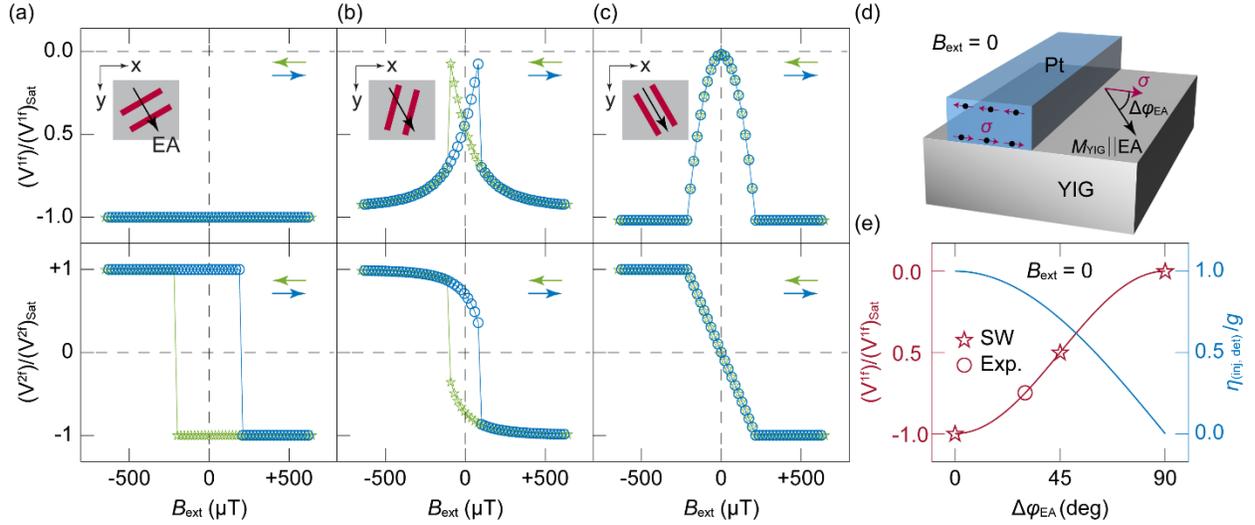

**Figure 4: Utility of Magnetic Anisotropy For Field-Free Magnon Transport.** (a-c) Macrospin simulation of hysteresis loops of $V^{1f}$ (top) and $V^{2f}$ (bottom) signals for three device configurations with different twist angles between Pt wires and the easy axis. In the inset, the red lines represent Pt wires, and the arrow marks the direction of the magnetic easy axis. In all device configurations, $B_{ext}$ is applied in-plane and perpendicular to the Pt wires. (d) Definition of the twist angle $\Delta\varphi_{EA}$, the angle between the polarization ($\sigma$) of the spin accumulation in Pt and the magnetic easy axis. Twist angles of $\Delta\varphi_{EA} = 0°$, $45°$, and $90°$ are assumed for configurations shown in panels (a)-(c), respectively. (e) Modulation of the interfacial coupling efficiency (right axis) and, by extension, the remnant value of the $V^{1f}$ voltage (left axis) with the twist angle. The coupling efficiency assumes an analytical form of $\cos(\Delta\varphi_{EA})$, which leads to the $\cos^2(\Delta\varphi_{EA})$ dependence for the first harmonic signal. Three pentagon symbols represent the zero-field values of $V^{1f}$ obtained from the macrospin simulation of devices shown in panels (a)-(c). The circle shows the experimental Data obtained from our device ($\Delta\varphi_{EA} = 30°$). Adjust main text for the experimental data point on Panel (e)
24

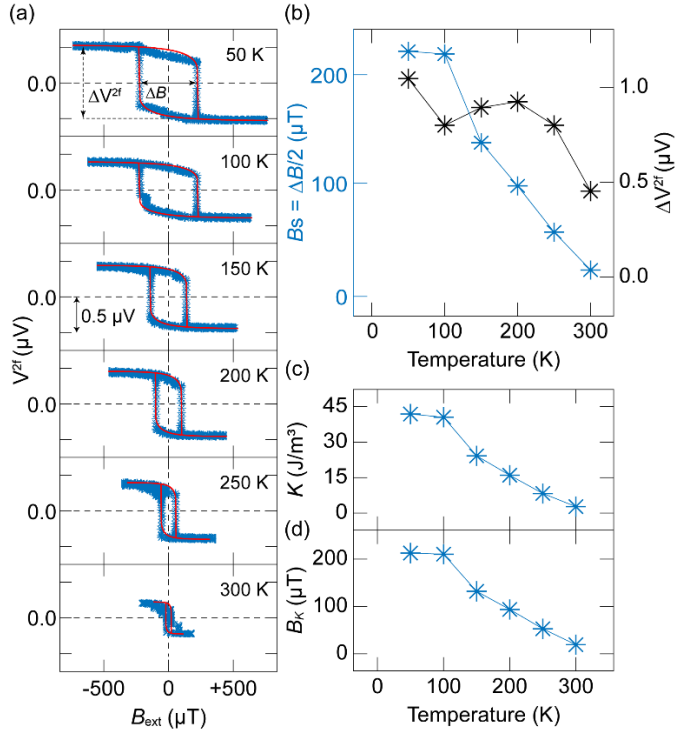

**Figure 5: Temperature Dependence of Magnetic Anisotropy.** (a) Temperature dependence of the hysteresis loop in $V^{2f}$ voltage. The height ($\Delta V^{2f}$) and width ($\Delta B$) of the hysteresis loops are indicated on the top panel. Red curves represent macrospin simulation fitted to the experimental data. $B_{ext}$ is applied perpendicular to the Pt wires at $\varphi_B = 0°$. (b) Evolution of $\Delta V^{2f}$ (right axis) and the switching field $B_s = \Delta B/2$ (left axis) with temperature. (c, d) Anisotropy energy and the effective anisotropy field at different temperatures. These values are obtained by fitting the SW model to the experimental data, while accounting for the temperature dependence of the YIG's saturation magnetization. A device with the injector – detector distance of ~0.5µm is used here.